# Random and Coherence Noise Attenuation Complete Sequence for 2-D Land Seismic Data Acquired in Iraq

Ahmed J. R. Al-Heety [1*], Hassan A. Thabit [1]

[1] Dep. of Seismic Data Processing, Oil Exploration Company, Ministry of Oil (OEC), Baghdad, Iraq

[*] Corresponding author e-mail: ahmedalheety@gmail.com

**Abstract:** Noises are common events in seismic reflection data that have very striking features in seismograms, affecting seismic data processing and interpretation. Noise attenuation is an essential phase in seismic processing data, usually resulting in seismic interpretation improvement that enhances the signal to noise ratio. Groundroll presence is the major fashion of significant noise in land seismic survey. It is type of coherent noise present in seismograms that appears as linear events, in most cases overlapping the reflections and probably making it challenging to recognize. There are several domains used in noise attenuation, Domain transformations is a complex algorithms standard tool used commonly used during processing of seismic data and imaging processing, So a large number methods have been developed to attenuate these types of noise. In the time-offset domain, the noise wave such as Groundroll and random noises, overlap each other over time; a different domain makes it easier to successfully isolate coherent, random noise and reflection events. Five steps are introduced to attenuate coherent and random noise, these steps are: FDNAT, AGORA, RADMX, SCFIL, DDMED as well as Time-Variant band-pass Filter. The results indicate that the different domains can actually reveal features and geological structures that have been masked by the noises present in current data. because encourage significant improvements in the final image quality in 2-D seismic section, so, these filtering techniques possibly give advantage to interpreter in particularly in structural and stratigraphic interpretation during the work of interpretation, especially in the exploration and characterizing possible traps.

**1. Introduction**

The seismic exploration technique is the most frequently used and well-known geophysical method between several methods of geophysical prospecting. The seismic data can be processed to expose details information of geological structures on scales from the dozens of meters of the crust to central core of the earth (Yilmaz2001; Kearey, 2002). Major part of most of its continued success is that raw seismic data is processed to yield accurate subsurface images of the geological structures. Oil exploration and production companies use a several methods to evaluate, attempt to recognize hydrocarbon reservoirs so the processing of seismic data becomes significantly important. Several elements can effect on accuracy of seismic data to initially ensure successful drilling and later to contribute such a clearer understanding of the characteristics of the reservoir. Appropriate and acceptable noise attenuation strategies help increase the potential advantage and significant impact of seismic data in exploration, production and development. Advancements in signal processing methodologies have meaningfully impacted to the geological interpretation of data for interested area. Seismic data processing requires the applied several successive algebraic, statistical and signal processing methodologies, which are usually mixed with experienced and skilled geophysicist in specific interpretation. These seismic data processing stages include such as geometric spreading compensation, static correction, frequency/wavelength filtering, velocity analysis, deconvolution, time/depth migration, etc. (Yilmaz, 2001). After all, the processed data delivered to the interpretation analysis, which principally targets to generate a simple, credible geological represented model that is fully corresponding with the measured data. The noise in seismic exploration specifically implies to an uninterpretable or undesired portion of recorded seismic signals due to a several of reasons. These undesirable events may probably be considered as signals, but generally these undesirable noise events give inadequate or Confusion info about the subsurface and are usually referred to as random noise and coherent noise. The Fourier transform is absolutely essential to seismic data processing because it usually provides many mathematical theorems that are valuable in seismic image analysis. It therefore generally applies to almost all processing phases. Seismic signals are instances of seismic waves that are strongly considered and processed in different domain. The Spectrum analysis is used when denoting to an image's frequency or wavenumber content (Al-Shuhail, et al., 2017). A seismic trace is describing a seismic wave in a receiver position. The seismic trace in the digital form expressed by a time series, that are could be fully described as a discrete summation of a number of sinusoids each with a single peak amplitude, frequency, and phase-lag (relative alignment). The forward Fourier transformation solves the analysis of a seismic trace into its sinusoidal components. On the other hand, the inverse Fourier transformation solves the synthesis of a seismic trace from the discrete sinusoidal components. The first step in noise attenuation is analyzing seismic data to identify the sources and physical characteristics of noise regardless of the noise source, the characteristics usually fall into two categories: the first is Coherent Noise (Ground Roll, Guided Waves, Multiples and Power Line) while The second is random noise (e.g. spikes) (Carolyn, 2010). To be able to provide perfect subsurface images, we must overcome the issues of noise. Groundroll is identified by high amplitude, low frequency and group velocity. Receiver arrays are used in the field to significantly reduce Groundroll. Because of lateral inhomogeneities in the shallow layers, the groundroll may have sophisticated backscattered components. Guided waves are constant, particularly in shallow water depths (marine) records with rigid seabed; furthermore these waves are probably developed in land records. The dispersive property of these waves makes it comfortable to clearly distinguish them from shot records. These types of seismic waves are mostly suppressed by stacking Common mid-point. In principle, due to their prominently linear movement, they can also be actively suppressed by dip filtering strategies based on two dimensional Fourier transformations of shot records.

## 2. Materials and Methods

The sector of processing seismic data is huge. Published contributions (books, articles, reports) go back to the early beginnings of seismology of reflection in the 1920s and continue to accelerate with the early introduction of digital computers in the Sixties decade (Yilmaz, 2002). Seismic data mainly involves of a signal and noise components. The general definition, of noise is any recorded waves that interfere with the wanted signal. The variability of noise types sometimes certainly makes separating of signal and noise a problematic and not easy procedure. However, the effective noise suppression is essential for high resolution imaging. Subtracting the noise from seismic data is a significant step towards super confident interpretations. The seismic data could contain both random and coherent noise. Each one of these should be resolved in a special way and can require additional application of much more than one strategy to produce optimum results. Attenuating high-amplitude noises, like ground roll, is indeed a great challenge in the processing of seismic data. The last subsurface image may be offers some incorrect interpretation info without attenuating these high-amplitude noises (Guo and Lin, 2003). The signals are primary reflection and diffraction. Any other types of events are generally considered as noise. Surface waves, such as groundroll, are a main source of coherent noise in base-land seismic data, particularly in the current study area due to lateral inhomogeneity in thinner-surface layers and variable low-velocity layer (LVL), so all these layers belong to quaternary and recent deposits. The variation in weathered zones conditions therefore usually explains into variance in the noise character. Guided waves are produced by specific geological conditions on the subsurface, such as layers of high-low-high acoustic impedance. Guided waves are recognized by high velocity and amplitude. The dispersive property of these surface waves (such as Rayleigh and love waves) gives somebody comfortable to distinguish on shot records, as well as make up speedy arrivals. Coherent noise such as Groundroll can be measured at a specific velocity or velocity group. The noise can then be attenuated by the velocity depending on the data. F-k filters have generally been used for this type of noise removal. In practice, the SNR is not simple to quantify due to difficulties in separating signal and noise. The raw seismic data haven't similarity to the structure inside the earth. It is mostly an illustration of the experimental info of how the data was acquired. Seismic data processing uses full knowledge of wave propagation and acquisition geometry to create subsurface geologically meaningful images. Seismic data processing consists of five groups of regulations and corrections: time, frequency-phase content, stacking, and positioning (time /depth migration) data. These procedures increase SNR, and correct data for many physical processes that confusing the needed (geological) seismic data info, and reducing the size of data to be evaluated by the geophysicist (Iqbal et al., 2018). The seismic data provide the geological information about the shape and the comparable destinations of the geological structure and lithology. In areas of good data quality, lithology estimates can be produced based on velocity information. The pores constituents could even be predicted from the amplitudes of reflection due to the abnormal amplitude (anomalies) generated of gas concentrations. Recognizing of structures shape as a function depth basically allows petroleum company explorers to give possibilities in the survey area to discover commercially exploitable hydrocarbons. The velocities of the seismic waves can be extracted from seismic data or computed in wells and used to transform known reflection times into estimated reflector depth. In seismic processing data, there is no single processing algorithm that can remove all noise types. Nearly all denoising methods have the common methodology of easily transferring the data to any domain where the signal and noise component can be separated. Practically all denoising methods have the common methodology of easily transferring the data to any domain where the signal and noise component can be separated. The recognized noise is therefore removed before the data component is transformed back into a typical (t –x) domain. Therefore, the challenge is to find a domain in which noise and signal are well detached (Elboth et al., 2008). Below we will briefly explain the step-by-step data processing sequence for noise attenuation that is directly relevant to a 2-D seismic data set that will come later in this article. Figure 1 show the sequential noise attenuation flowchart used in this study.

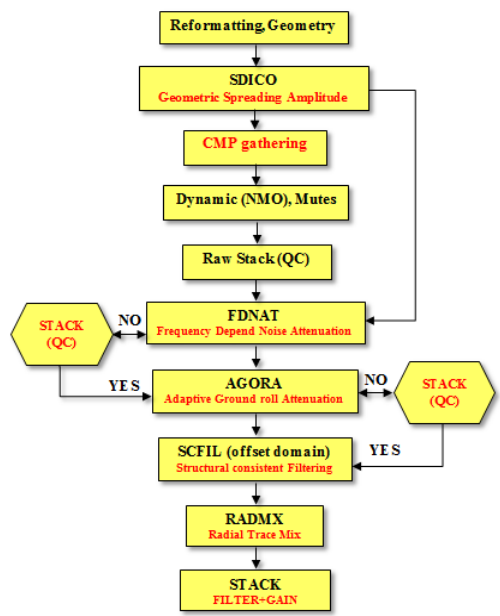

Figure 1 Sequential noise attenuation Processing Flowchart used in this study.

1. **Reformatting and Geometry up-data**

The Input data was supplied from OEC data bank in SEG-D format. This format was transcribed from SEGD to the internal Geovation data format (SDS) for the data geometry update. The geometry builds–up by using onset (CGG application). We are Identify the information about the acquisition geometry needed to process seismic data. Define the term Common Mid-Point (CMP), Define the term nominal fold, Describe the relationship between receiver spacing and CMP spacing, record length, sample interval, Shot interval, Trace interval, spreading type, first and last shot number…etc. these parameters did and exported with ETQXY module, This module is designed to update certain header attributes which are filled with values from a 2-D or 3-D land data acquisition. Consequently, the Input traces are assumed to correspond to an acquisition strip or swath and to be ranked by recording number, after having been labeled by module SEGIN. For both the 2-D and 3-D options, the trace header must contain the record number and the channel number, shot point and receiver coordinates, geometry information at the end all parameter built-in in information file called UDFILE. Then we have done QC for the Shot Records by checking start time curves above shots. The start times calculated by the equation: [(OFFSET_NB/replacement velocity)*1000] and build in Header (USER_STARTTIME), Also, we are preparing Field statics (in this study used up-hole static) calculation to use it in the next steps such as Stacks (QC) and AGORA module. Finally, seismic data is merged with field geometry file.

2. **Spherical divergence (Geometric Spreading) correction (SDICO)**

A field record describes a wavefield produced by an individual shot. Theoretically, individual shot is considered as a point source that yields a spherical wavefield. The earth has two impacts on a propagating wavefield: (1) in homogeneous geology (medium), energy intensity decreases proportionately to the square wavefront radius. The amplitude of the wave is proportional to the square root of energy intensity; it decays as 1/r. In practice, velocity generally increases with depth, which simply causes further wavefront divergence and significantly fast decay amplitudes with distance. (2) The frequency components of the original source signal varying with time as it propagates. In particular, the high frequencies are absorbed earlier than low frequencies. This is due to the intrinsic attenuation in rocks. Therefore, to bring up any signal that may be present in the deep portion of the record, this earth effect must be removed. By using the primary velocity function to correct geometric spreading, the amplitudes of the dispersive coherent noise and multiples have been overcorrected (Yilmaz, 2001). The factor (1/r) that explains the decay of wave amplitudes as a function of the spherical wavefront radius is relevant for a homogeneous medium without attenuation. Amplitude decay can be described approximately by 1/[v2(t) t] for a layered earth (Newman, 1973). Where (t) is the two-way travel time and V (t) RMS velocity of the primary reflections (those reflected only once) averaged over a survey area. The function gain of geometric spreading correction express by:

$$g(t) = \frac{v^2(t)t}{v_\circ^2 t_\circ} -------(1)$$

Where (v0) refers to the reference velocity at a definite time (t0)

SDICO module applies or removes compensation for the effects of geometrical spreading calculated using Newman's formula. The amplitude correction process was utilized to seismic data in order to compensate for the amplitude decay which developing from the propagation of the seismic wave from a point source in a layered medium. For optimum amplitude compensation, SDICO CGG programs were applied for data. SDICO program applies a compensation for the effects of geometrical spreading computed using P. Newman's formulae. Each sample is multiplied by the followed equation:

$$T^n V^m (T\circ) --------(2)$$

Where (T) is the TWT of the sample in (seconds), (To) is the NMO corrected time, V (To) is the stack velocity at time To for the CDP being considered, (n & m) are the user-defined powers of T and V.

SDICO program has a supplementary term to take the offset into account during computation. Also, CDP ordering is recommended to optimize execution time (the velocity functions read CDP by CDP) (Geovation User's Manual, 2010). Figure 2 shot gather before and after applied Spherical divergence compensation (SDICO).

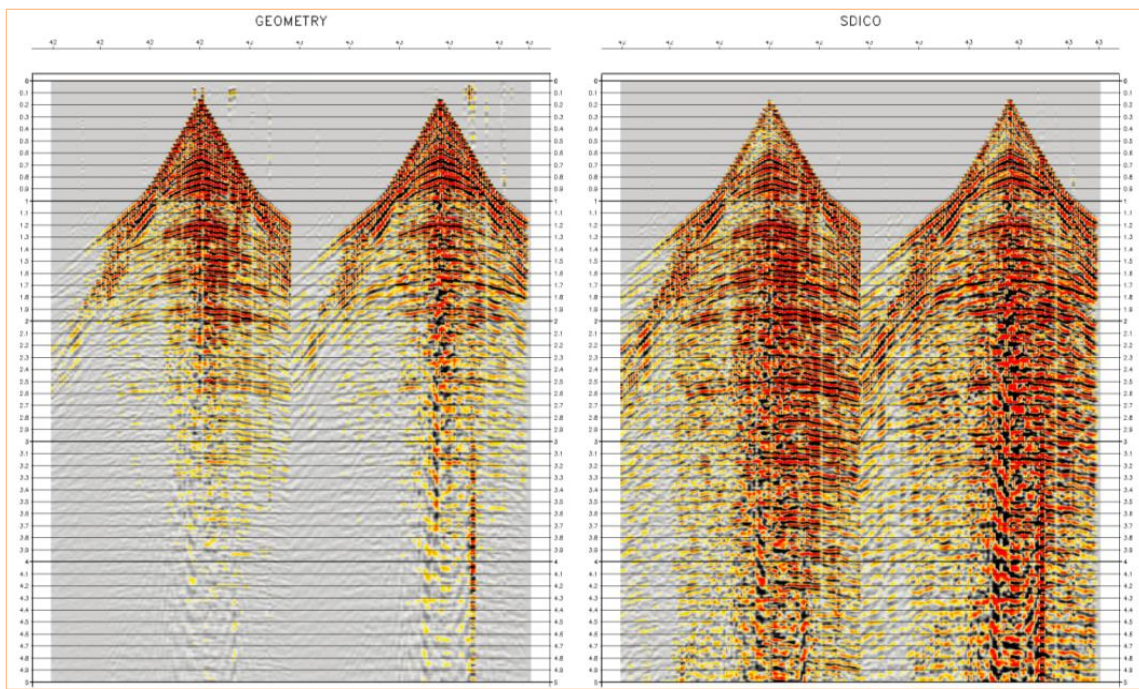

Figure 2 Shot gather before and after Spherical divergence compensation (SDICO)

## 3. Frequency-Dependent Noise Attenuation (FDNAT)

In seismic reflection data, random noise can be generated from several sources, such as poorly fixed geophones, wind motion, human activities, traffic, and transient movements neighboring recording cable or electrical power noise. Some random noise invariably exhibits spike-like property (Yilmaz, 1987); also, random noise also can be generated by scattering from near-surface anomalies such as gravel, boulders, or vuggy limestone (Dobrin and Savit, 1988). Although stacking can at least relatively attenuate random noise in shot gathering data, remaining random noise after stacking will significantly reduce the accuracy of final data interpretation (Yilmaz, 1987).

This step removes unwanted noise (Anomalous Amplitude) while preserving the signal and enhancing the seismic image. Many sources of noise contaminate seismic recordings, distorting image quality, troubling interpretation, and distorting the seismic signal prior to inversion and other reservoir characterization stages. The suppression of noise that can be produced from shallow subsurface is a challenging objective to accomplish in land seismic processing. On the other hand, some properties can be obtained from input data to feed an adaptive filtering to eliminate "pseudo-random" or coherent noise. The method is an adaptive automated frequency dependent amplitude threshold method to distinguish and remove high amplitude "spiky" and some of coherent noise. The FDNAT module attenuates high-amplitude noise in decomposed frequency bands. It uses frequency-dependent and time-variant amplitude threshold values in defined trace neighborhoods to detect and suppress noise specific to different frequency ranges and different times. There are three options are available in FDNAT. These options (Blank option, SO option and GC option) which differ in the way they define strength comparison neighborhoods for noise identification. The first one Blank option is defined trace neighborhoods by trace gathering (one or more groups) and trace distance within the gathering. The second one SO option defines the comparison neighborhoods entirely based on the trace input order (not deal with gathers). The size of the neighborhoods is given by a number of traces in the neighborhood. The neighborhood for an input trace generally consists of (NC/2) traces input before, and (NC/2) traces input after, the current trace. The parameter NC only works for SO option. The third one GC option, here we can say that both the blank option and the SO option have their advantages and shortcomings. So, Option GC is designed to combine these two options to create new ways of defining the comparison neighborhoods. (Geovation User's manual, 2010). Figure 3 illustrates the definition of the strength of a trace sample. The strength of a sample is defined as the average of the absolute amplitude values of all samples in a time window centered at the given trace sample and shows the method used for interpolation of threshold (TH) values as a function of frequency and time.

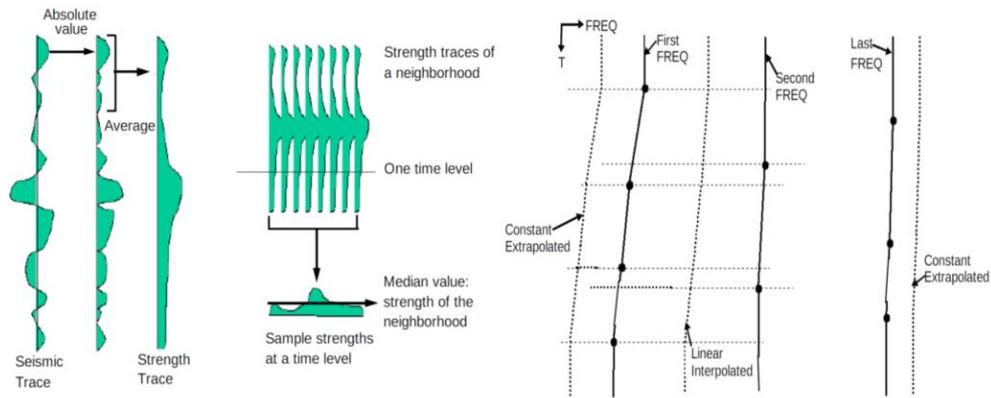

**Figure 3** Illustrates the definition of the strength of a trace sample (left); (Right) show the method used for interpolation of threshold (TH) values as a function of time and frequency.

A median filtering was applied in different frequency ranges (module FDNAT). The module was applied in two modifications: an entire CDP gather (harsh filtering) and only neighbor traces in the window of 41 traces (FDNAT Blank modification, soft filtering) was used as a group by which an average value is computed. Table1 summarizing the applied Parameters. Figure 4 shows the result of Shots data before, after and difference applied module (FDNAT).

Table 1 Application Parameters (FDNAT) for shot gather

| **Frequency (HZ)** | **Time(T)** | **Threshold value(TH)** |
|---|---|---|
| 10, 20,40,60 | 0 | 8 |
| | 1000 | 6 |
| | 2000 | 2 |
| | 5000 | 1 |

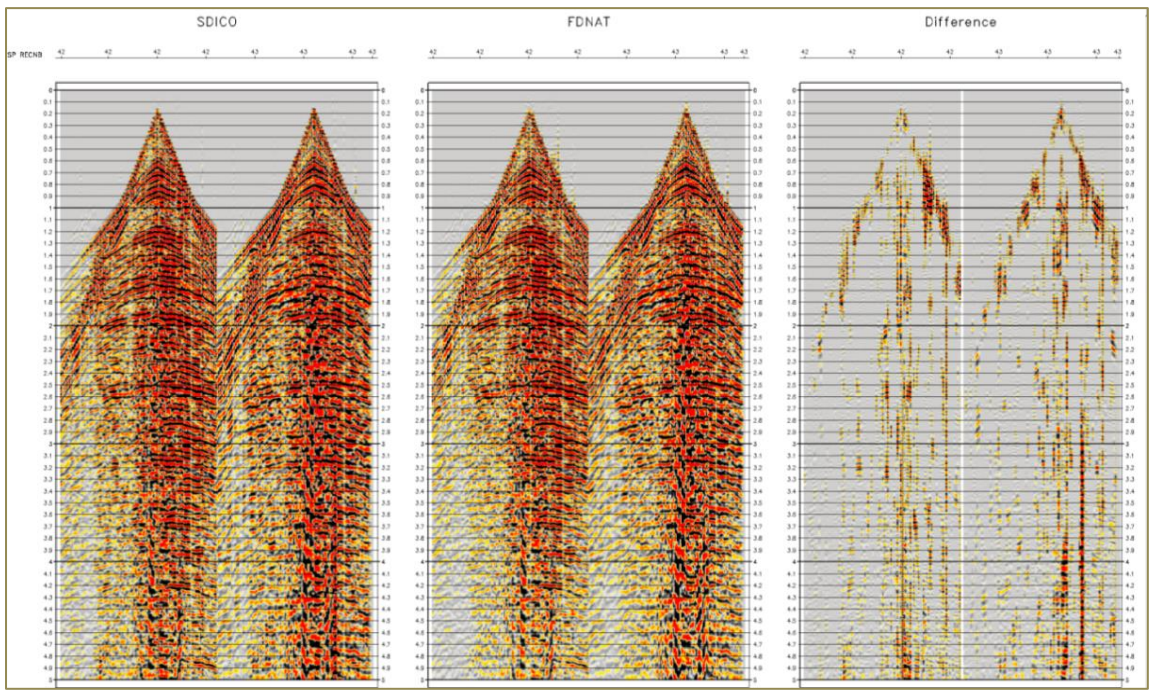

Figure 4 Shots gather before, after and difference applied module (FDNAT).

## 4. Adaptive Ground Roll Attenuation (AGORA)

The effect of near-surface variations is the main reason for modest quality of seismic data in the land seismic survey. Groundroll suppression is the initial problems that really must be adequately adopted throughout data processing steps. groundroll is surface wave which recorded on a vertical geophone as "pseudo-Rayleigh" waves. They are direct result of interfering primary and vertical shear waves (SV) that travel along or near-surface. Groundroll arrives at once from the seismic source hence, for inline cables (of 3-D data) it is a linear on near offsets but seems hyperbolic on the near offsets. Groundroll can be effectively dispersive and aliased and behave as guided waves, meaning that there are different phase velocities for every frequency. Over the past three decades, various solutions have been established to attempt to suppression the Groundroll in 2-D and 3-D seismic data involving *f-k* method, Wavelet Transforms, High-Resolution Radon, and Elastic Modeling solutions. All of these techniques at the present-day applied sometimes alone and sometimes in cascaded applications. However, these techniques generally have specified standard when used over large surface areas and therefore have a very low response to the changing physical nature of the near-surface layer. This sometimes results in the absence of coherent artifacts and leads to low protection of primary amplitudes caused by sever or inadequate filtering (David. et al, 2008).

In this step, we will describe methodology that conducts adaptive-filtering of aliased and dispersive surface waves at their real spatial coordinates called (AGORA). The principal AGORA module is modelling signal and noise in the *f-x* domain in each gather by separate characteristics contained them.

The signal is modelled as hyperbolic events, whose trajectories are expressed by (rms) velocities. The ground roll and guided waves are modelled as a series of dispersive linear events, each characterized by phase and group velocities. This modelling utilization the true distance between sources and receiver, i.e., the spatial sampling can be irregular. A least-squares iterative method is then used to adapt this model to the input data before subtraction of the surface noise. Note that this outline is efficient if the current frequency is reasonably close to a defined central frequency. The principle of the modelling is explained in detail in Perkins and Zwaan (2000) and Le Meur et al. (2008a). The AGORA process can be described by the following main stages (Geovation User's manual, 2010):

1. Extractions of the ground roll characteristics via a frequency-velocity phase diagram.
2. Wavelet domain. Wavelet filter banks allow a multi-resolution approach with a split of the input data in several frequency-wave number sub-panels using a "highly" reversible wavelet transform.
3. Modeling in the (f-x) domain of aliased and dispersive surface waves is done for each sub-panel using the most adapted set of parameters derived from the data itself via the frequency-phase velocity panel. Adapt Groundroll model to fit data and subtract Groundroll model from data.

The basic is that the input (raw) data is a mixture of signal in addition to coherent and random noise. The signal (S) is modeled as hyperbolic events whose trajectories are described by stacking velocity using the equation (Le Meur et al., 2008b):

$$S^{j,k} = \exp\left[if\left(\sqrt{t_j^2 + \frac{xk^2}{Vrmsj^2}}\right)\right] - - - - - (2)$$

The coherent noise such as Groundroll is modeled as a series of dispersive linear events, each characterized via phase and group velocities using the following equation:

$$GR^{j,k} = \exp\left[i\left(\frac{f0}{Vp_j} + \frac{f-f0}{Vgj^2}\right)xk\right] - - - - - (3)$$

For a jth event: (to)zero offset travel time, (*xk*) true shot to receiver distance, (f0) central frequency of the wave, (vpj) phase velocity, (vgj) group velocities. These events form the components of the matrix A with column and row indices j and k.

In the frequency domain, the input (raw) data is represented by a matrix (D) which can be defined by a matrix (A) which present the dispersive linear and hyperbolic events multiplied by a vector (W) which comprising an unknown wavelet corresponding to the signal and groundroll adding to a proportion of random noise( N) as explained by the equation:

$$D = A.W + N --------(4)$$

Rewriting equation (5) in suitable terms, the least square iterative inversion strategy is used to separate the groundroll from the raw data (input). Notice the majority of the cases where the groundroll may have a broadband of more than 30 Hz, however, it is resolved by splitting the data into multiple frequency bands that allow multiple different central frequencies to be used to optimize coherent noise modeling (David. et al, 2008). Coherent linear events in the (t-x) domain can be separated by dips in the (f-k) domain. This simply enables us to reduce certain types of undesired energy from the data. In particular, consistent linear noise in the form of a ground roll, guided waves, and side-scattered energy usually convoluted primary reflections in recorded data. These types of noise are generally separated from the (f-k) domain's reflective energy. The results are shown in Figure 5 and Figure 6 The strength of this method is that it is data driven rather than deterministic. It makes less assumptions than traditional methods, e.g. it uses the real (irregular) source/receiver positions rather than always assuming regular spacing.

Noise characteristics (group and phase velocity) are obtained for each input group instead of using fixed parameterization, e.g. dip cutoffs in conventional f-k filters. The adaptive subtraction reacts to variations in amplitude to help prevent primary damage. Finally, this method is flexible and adaptable and can be applied in a number of different domains to best suit the workflow or data needs, due to acquisition geometry. Figure 5 show the groundroll property over raw shot

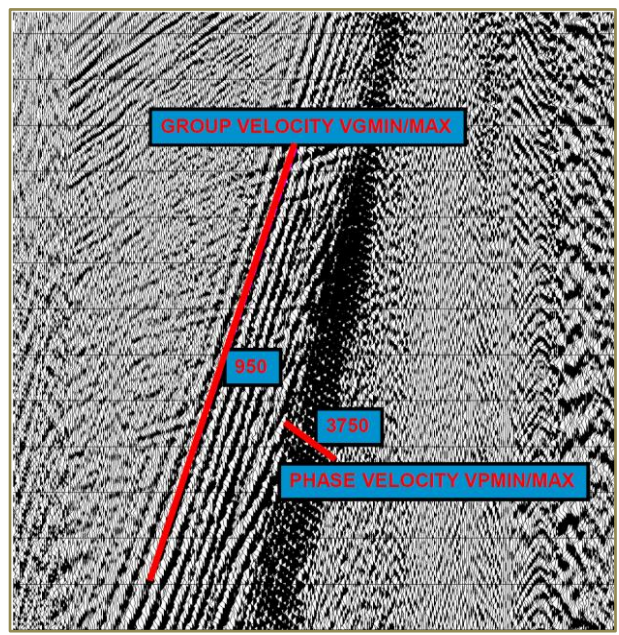

Figure 5 groundroll property over zoomed raw shot

Table 2 Application Parameters (AGORA) for shot gathers

| Parameter | Value | Parameter | Value |
|---|---|---|---|
| Min Frequency/FMIN | 2 | Max Frequency (FMAX) | 20 |
| Min. GR velocity/ VGMIN | 300 | Max. GR velocity (VGMAX) | 950 |
| Min. phase velocity/VPMIN | 500 | Max. phase velocity (VPMAX) | 3750 |
| DBST(HISTA) | Static | % of Nyquist wave num. to preserve/KHCUT | 100% |
| Num. of linear event for noise/NUMMOD (parameter for *f-x* modeling) | | | 4 |

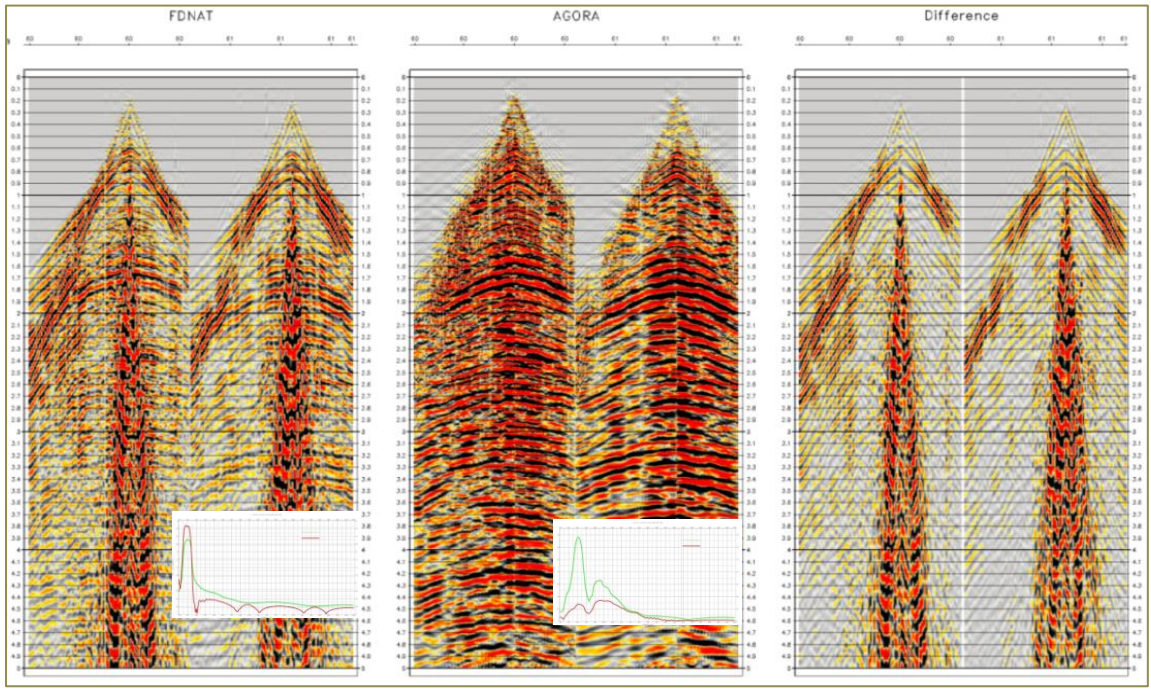

Figure 6 Shots gather before, after and difference by using module (AGORA).

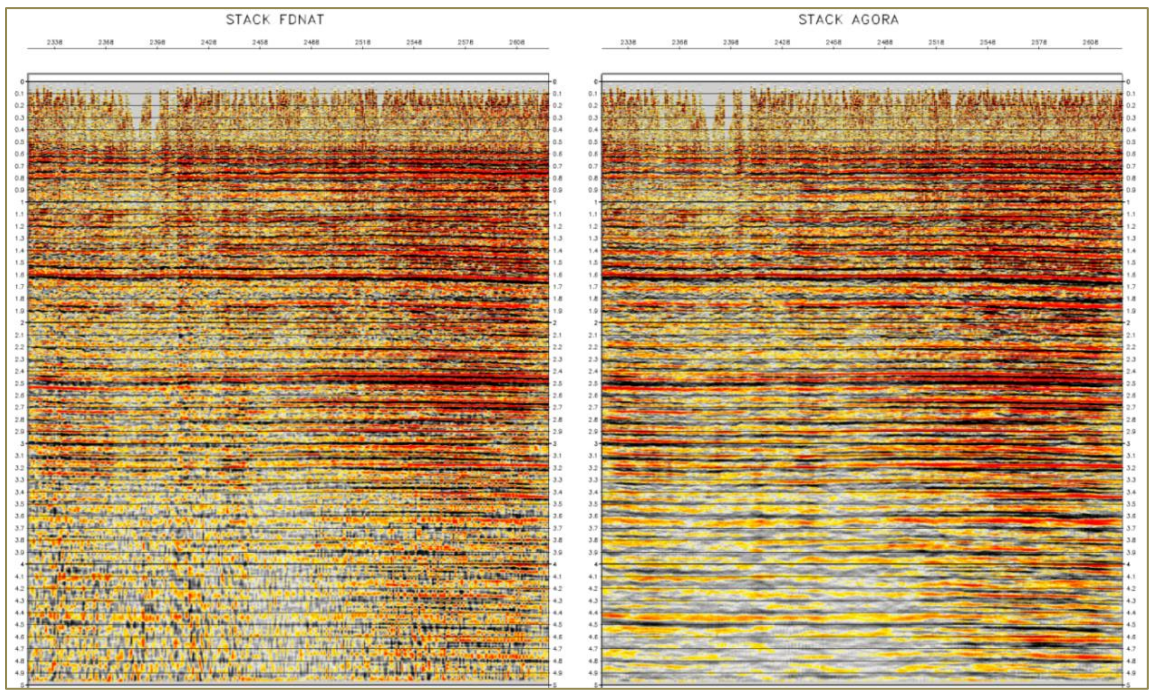

**Figure 7** Show stacks data before and after using module (AGORA).

## 5. Radial Trace Mix for Noise Attenuation (RADMX)

Henley (1999; 2000; 2003) introduced radial trace domain(R-T) methods to attenuate coherent noise in seismic data, partly based on previous perform by Claerbout (1975 and 1983), who presented radial trace transformation mainly for apply in migration and associated imaging processes. radial trace domain is a coherent noise suppression methods take advantage of the fact that linear noise fragmented from reflection events can be accomplished in the (R-T) domain by aligning the transform coordinate trajectories with the coherent noise wavefronts in the (*t-x*) domain. As a result, linear noises projected across several constant offset traces of (*t-x*) gather are collapsed into relatively narrow groups of constant-velocity traces in the R-T domain. In addition, the apparent frequencies of these reflection events decay dramatically, often to the sub-seismic range (Henley, 1999). The Radial Trace Mix for Noise Attenuation (RADMX) module is designed for Random, coherent and slanted noise attenuation in common offset sorting (offset-class). It stacks each trace of a gather with other neighboring traces falling within a user-specified radius to create a noise reduced version of the same trace. Traces going into stack are weighted according to their radial distances (calculated from the position of the trace being processed). The module provides different weighting schemes. Here we are used Constant weighting scheme. Although this module has been designed to work on common receiver gathers (receiver domain), any type of gather domain can be used. RADMX designed to perform noise attenuation with assuming traces have NMO applied, stacking traces locally (offset domain) around each trace produces a noise suppressed version of the trace. RADMX has a built-in differential move-out application where the user defines a velocity files so that the program can apply the differential move out internally. Alternatively, NMO can be applied externally using any other NMO application module, in which case RADMX should be run without specifying a velocity file and then the NMO should be removed. Spatial substacking based on 7 traces (module RADMX) was used for additional random and slanted noise attenuation by offset classes. We are using the flowing parameter in RADMX module:

- RADIUS= (200 in distance units) within which the neighboring traces of the input Limits.
- YMX= Maximum fold. FMX= Maximum fold expected to be found within the area of stacking.
- ATTX & ATTY= Attribute for the x-coordinate and y-coordinate respectively of the input trace over which the neighborhood will be checked).

The weighting scheme (Constant weighting) working on all non-zero time samples to weighted them equally. In other words, if the number of non-zero time samples (on a given time slice) is N, then the weight is 1/N for each sample. The neighborhood of each input trace is checked using several parameters, such as RADIUS. Traces falling in this neighborhood are weighted according to their radial distances from the input trace. Those weighted traces are then stacked to form one trace which is basically a noise-suppressed version of the original input trace. This process is repeated for each input trace until all the traces in a gather are processed. The radial mixing process that yields these noise reduced traces also seems to be an influential way for attenuating steeply dipping noise which otherwise leaks into stack volumes. The results of applied RADMX are shown in Figure 8 and Figure 9.

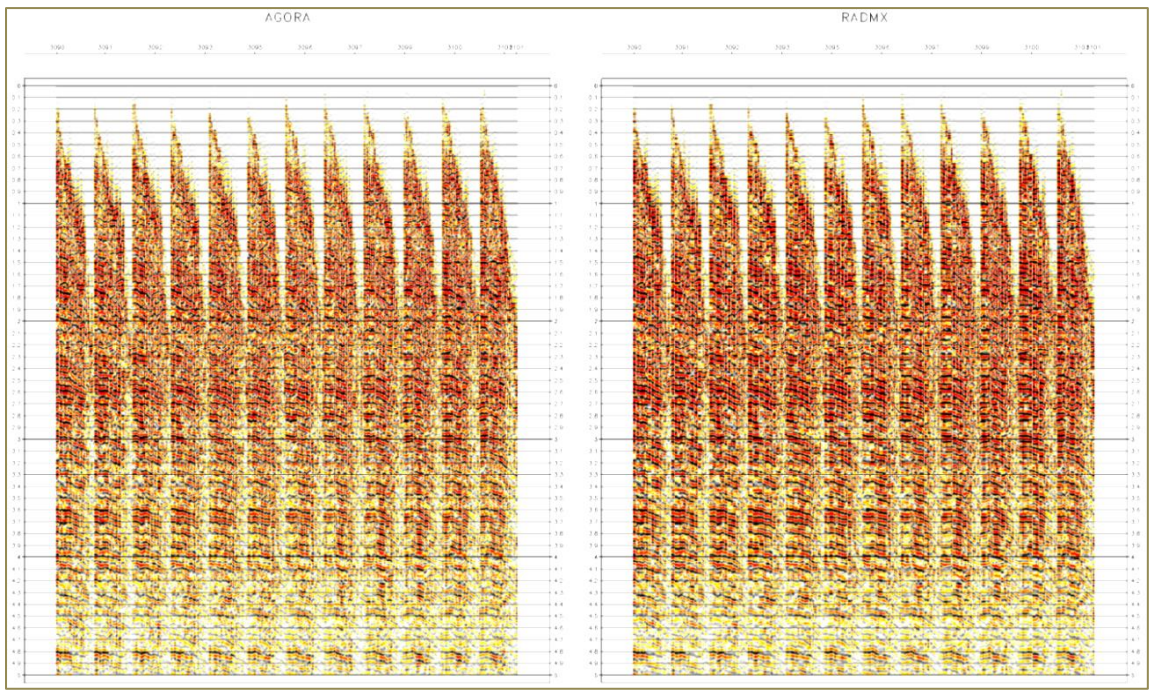

**Figure 8** Show CDP gather data before and after using module (RADMX).

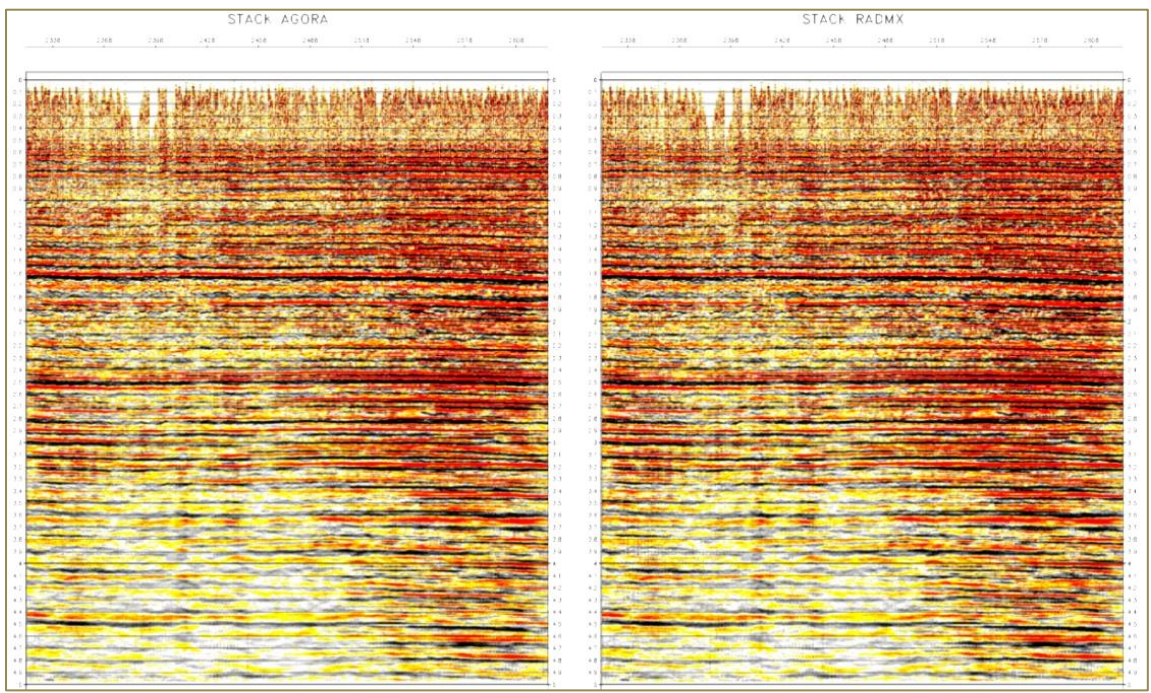

**Figure 9** Show stacks data before and after using module (RADMX).

### 6. Structurally Consistent Filtering (*f-x* filter) (SCFIL)

Structurally Consistent Filtering is random noise attenuation in offset domain which is an effective solution to execute 2-D dip filtering, driven by local, spatially varying dip fields. This current algorithm uses differential filters from dip steered composite, and tolerates filtering over large distances even when dips are geometrically slightly changing over short distances. Conventional de-noise processing usually involving geostatistical filtering (Le Meur et al., 2003), median filtering or low-pass Fourier filters (Siliqi et al., 2007) is quite effective. Such techniques, however, adjust the original spatial organization of the attributes along seismic horizons: a structurally consistent filtering method is therefore necessary. To avoid horizon picking, one can compute, from the seismic, the local dips is steered the filtering along the seismic layers. The first strategy class (Hoeber et al., 2006) mainly consists of applying non-linear filters such as median, trimmed mean, adaptive Gaussian filters, over planar surfaces, parallel to the dip. The planar assumption generally requires small, local application surfaces (Traonmilin and Herrmann, 2008). The first step of applying structurally consistent filtering is to obtain the data's structural information. We compute the corresponding dip field [p (t, x)]. This field will drive the structurally consistent filtering. There are several methods for computing such a field including scanning, plane wave destruction (Fomel, 2002), and common spectral analysis. Because we depend on the dip field to perform the filtering step, we have selected a least square strategy to ensure robust and stable dip estimation and then extract p from it. The second step is to filter once the dip field is calculated; then we have to build the filtering process. The solution we have chosen can be considered an adaptive convoluted *f-x* filtering. For each data point, a filter is calculated to select the known local dip. These local filters should have a property that the user can easily tune, the length of the filtering. This will allow the filtering strength to be defined in a natural fashion. The Structural criteria for seismic filters are that they should be smoothing homogeneous areas while preserving trends, edges and other details. There are many filters, and when running in a dip-consistent manner all these filters yield their maximum filtering efficiency and best structural conformity.

SCFIL module performs non-stationary filtering which attenuates random noise in common offset sorting. There are two methods: a time and space-variant low-pass filter and a structurally consistent filter. The filter is based on computed attributes. The main input is the data to be filtered. Dead or missing trace (zeroed trace) should be avoided if possible. The data is organized by gathers. An auxiliary input is used to compute the parameters of the filtering. It must have the same configuration as the main input. Traces must be at the same location (offset domain) as the main input. The output is the filtered data. Here we used structurally consistent filtering (SC) option; this option filters along the local dips computed on the reference data. To filter along the structure of the reference data we use local dips.

This dip a used to define local filters show in Figure10 which illustrate how to filter along the structure of the reference data. In our job we are use local dips. These dips are used to define local filters. The filtering along local structures defined by the stack was used for random noise attenuation in common offset sorting (module CSFIL, SC modification). A spatial window of 7 traces was used for filtering (filter length - 500 ms, frequency range - above 12 Hz). The results are shown in Figure 11 and Figure 12.

(**DX**) Trace spacing, in meters, (**YMX**) Maximum number of traces in a gather, (**LT**) Length in time of the analysis window, in milliseconds), (**LX**) Length in number of traces of the analysis window in the X direction, (**CL**)Correlation length of the filter, in meters, (**FMAX**) Maximum frequency to be filtered. so the Frequencies over FMAX will be zeroed (Default value: = Nyquist Frequency)., (**FBAND**) = Size of the frequency bands used to filter dips. (**TAPX**) = Num. of overlapping traces of the processing window. The taper is linear value: here we used (20).

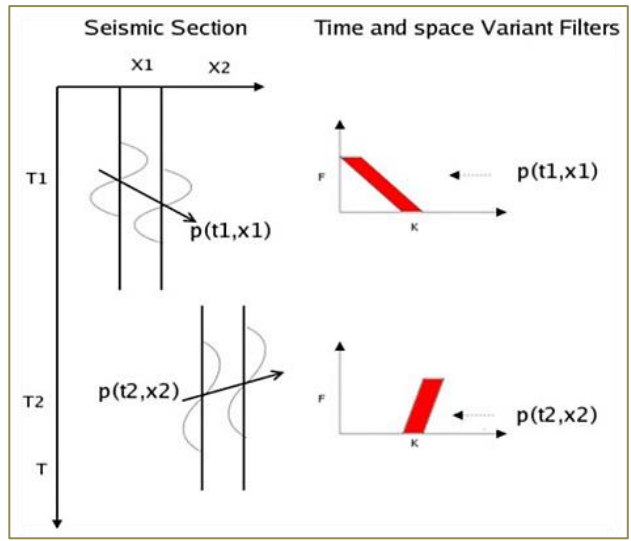

**Figure 10** shows how to filter along the structure of the reference data. We use local dips and these dips are used to define local filters.

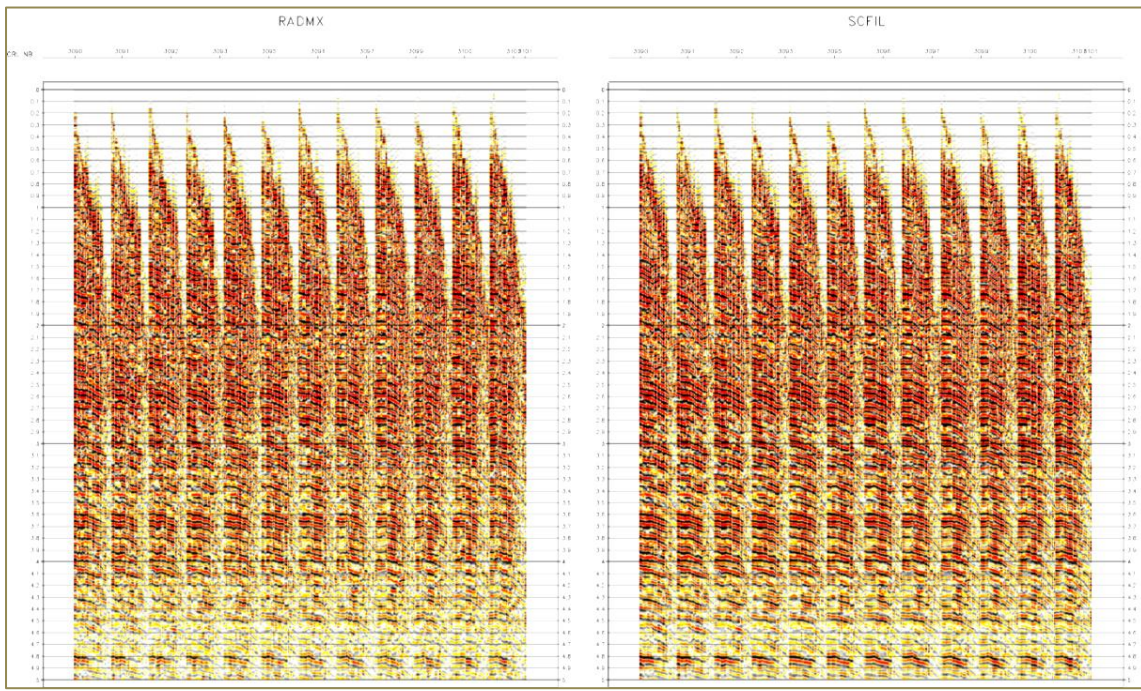

**Figure 11** Show CDP gather data before and after using module (SCFIL).

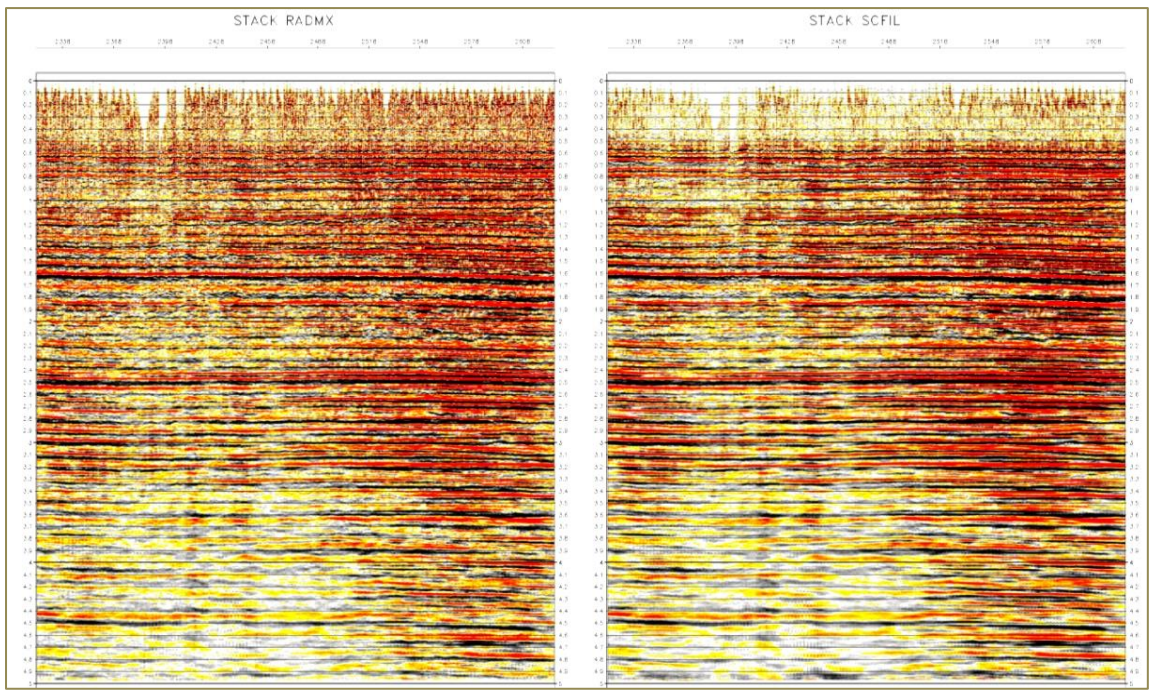

**Figure 12** stacks before and after applied (SCFIL) module.

7.  **Dip-Dependent Median/trimmed mean filtering (DDMED)**

This module applies a median or trimmed mean filter along the local dip of the data, thereby suppressing random noise and increasing trace-to-trace coherency. Dip-dependent median/trimmed mean filtering is a non-linear seismic noise attenuation process that can be applied to prestack or post-stack. It is a robust, effective and relatively easy operation that, in general, improves the signal/noise ratio and increases the trace-to-trace coherency of the input data. DDMED takes the input data and estimates the local (2D) dip at every sample point. A variable width median filter (i.e. a trimmed mean) is then applied along the dominant picked dip. At this point, the user can choose to apply a spatial weighting function as well. The filtered sample value then replaces the central sample value from the original local dip computation. Median filtering processes such as DDMED are generally better at removing local high amplitude noise events than other techniques such as F-K and F-X filtering. There are two key geophysical assumptions used in this module. The first assumption is the seismic data is coherent along a dominant local dip, it's implied that one specific dip can be picked, along which events are continuous. The second assumption that the amplitude values vary slowly along this dip. It's implied that, on a local scale, the median (or a spatially weighted trimmed mean) is a good estimate of the sample values. Given the above two properties, it is important to realize that any type of coherent noise along a specific dip (e.g. ground roll) will be enhanced in the same manner as real data (Reiter, et al., 1993; Holcombe and Wojslaw, 1992).DDMED carries out the following steps:

1. For every time sample, a dip search estimates the dominant local dip as shown in Figure 13.
2. A fraction of the maximum and minimum local sample values along this picked dip are removed from the filtering process.
3. A weighted average is computed from the remaining samples. The weights are either unity or a user-defined spatial weight function.
4. Finally, the original (central) time sample for the dip search is replaced by the filtered sample value.

In step (2), if all samples except the central amplitude value are removed, the process is equivalent to median filtering; and if no values are removed, the process is equivalent to taking the mean. While DDMED can be applied to data with variable trace spacing. Figure 14 Show the results of stacks before and after applied (DDMED) filter. Table 3 summarizing Applied Parameters for DDMED.

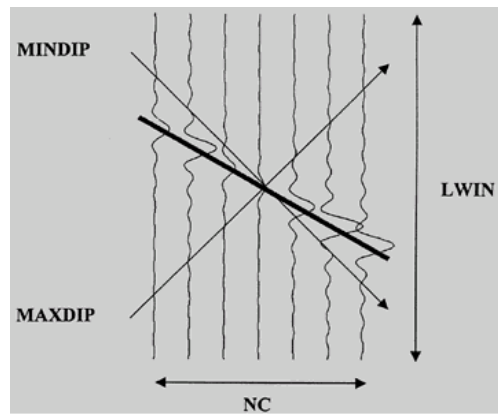

Figure 13 A ranges of dips is scanned about the central sample in the filter window. The dominant dip (denoted by the thick black line) is picked.

Table 3 Applied Parameters for DDMED

| parameter | Value |
| --- | --- |
| Size of the spatial window used in (dip search and filtering routines) (**NC**) | 3 (num. of traces in gather or stack) |
| Size of the time window (dip search) (**LWIN**) | 300 (ms) |
| The start time of the processing (**TI**) | 0 (ms) |
| **MAXDIP** &**MINDIP** (dip search) | 6 & -6 (ms/trace) |
| Num. of dips scanned in the (dip search) ND | 3 |
| Spatial half-width of a Gaussian curve (**SIGMA**) | 1 |

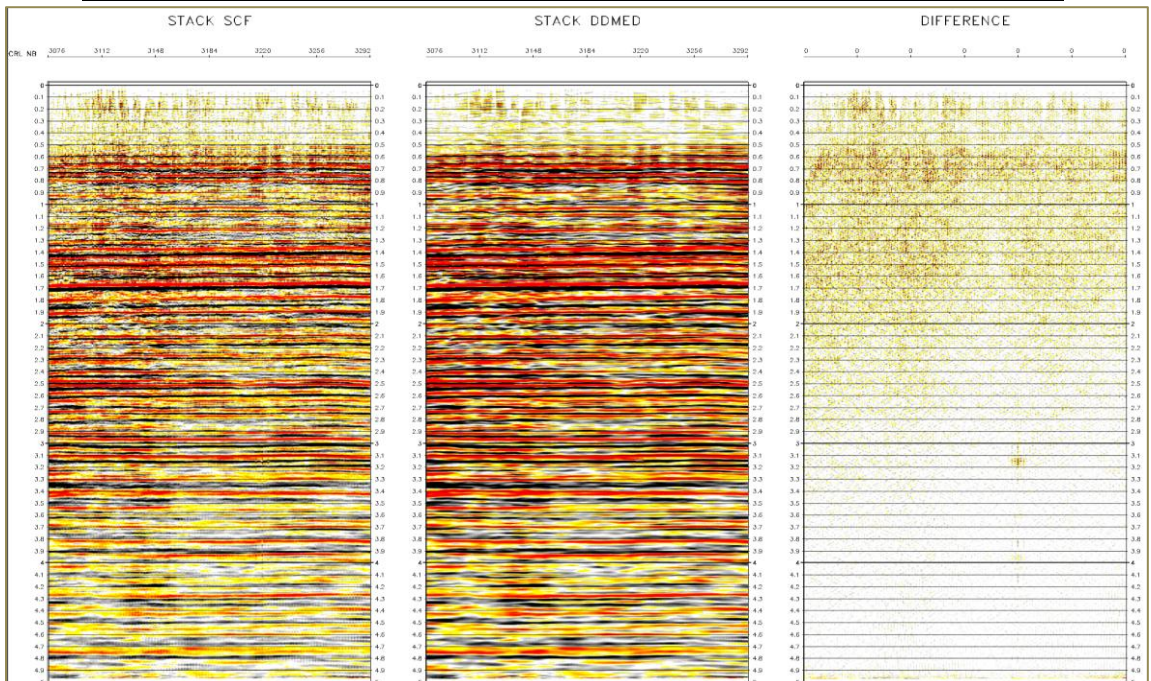

Figure 14 Show stacks before and after applied (DDMED) filter.

8. **Time Variant Filter (TVF)**

The frequency content of the seismic signal decreasing Vs. depth increasing because the higher frequency bands of the seismic signal are absorbed as the signal propagate deeper in time. Hence, the higher frequency of the signal is limited to the earlier part of the seismic section. Thus, vertical resolution is decreased significantly in the deeper part of seismic section. Because of these, noise in the stack or migrated section may still be broadband. For this reason, the seismic data processor looks for the optimum filter at variable times down the section which provides a good SNR. Therefore, we are need time-variant filtering (TVF). Also because SNR and signal spectrum changes with time and sometimes with space, it is more likely that one filter would not do for the entire section. Often a time-depended decay in SNR, thus, it is frequently necessary final processing step utilize TVF to uncover the clearest possible image of stratigraphic boundaries (Scheuer and Oldenburg, 1987). TVF usually are applied to stacked data. A constant bandwidth must be operated when filtering two sets of data that may have various vintages, source types, or noise levels. This is generally important when attempt to tie two lines and follow a reflector across them. The interpreter uses the frequency property of a sign horizon as a reference in the tracking process (Yilmaz, 2001). A series of filters were tested in discrete 5 Hz bands, from 25Hz to 70Hz, then a series of low-cut filters were tested (8Hz 10Hz, 12Hz, 14Hz, 16Hz, 18Hz,), in order to select the best time-variant filter for the final stack. Then a series of time-variant filters took place, to choose the best one. Tables (4) summarize Parameter of TV Filter. Figure 15 Show stacks data before and after applied (TVF). Table (5) summarizing of processing sequence for noise attenuation to current 2-D seismic data. Figure 15 and Figure 16 show the results of full processing sequence applied to current 2-D seismic data as while Figure 17 represents raw stack before and after noise attenuation.

**Table 4** Time Variant Filter Parameter

| Time (in ms) | Filter Band (in Hz) |
|---|---|
| 0-1100 | 14, 18-40, 50 Hz |
| 1400-5000 | 12, 16-35, 45Hz |

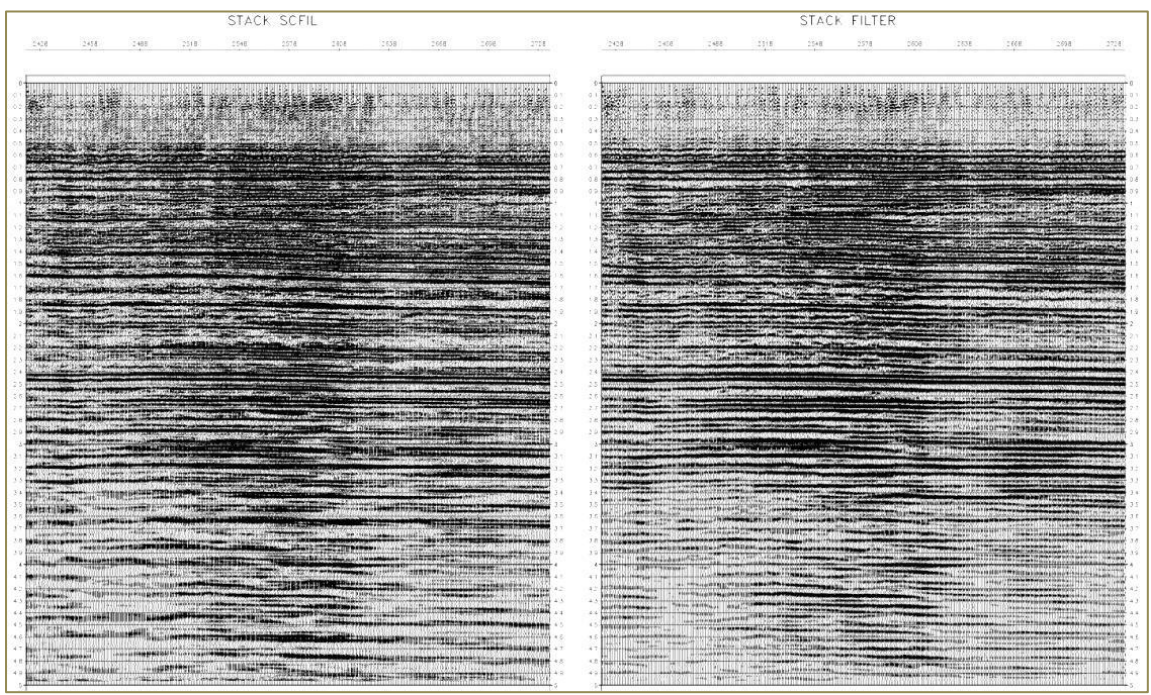

**Figure** 15 Stack data before (left) and after (right) applied (TVF) filter.

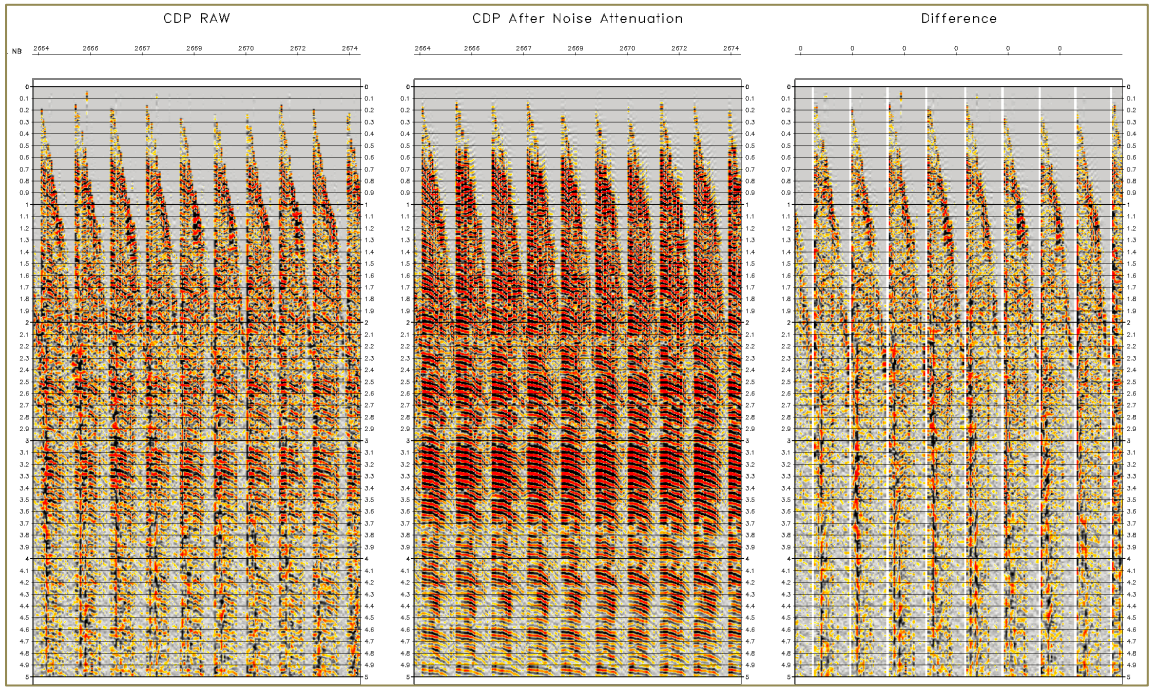

**Figure** 16 CDP gathers (A) before noise attenuation (B) after noise attenuation (C) difference

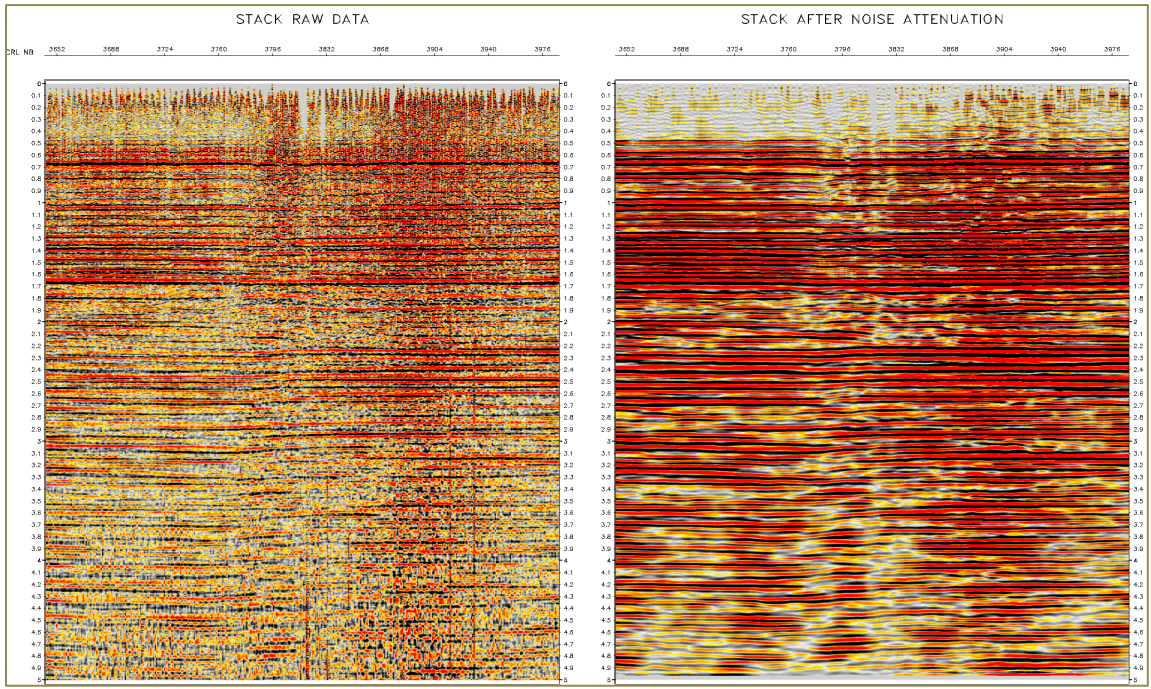

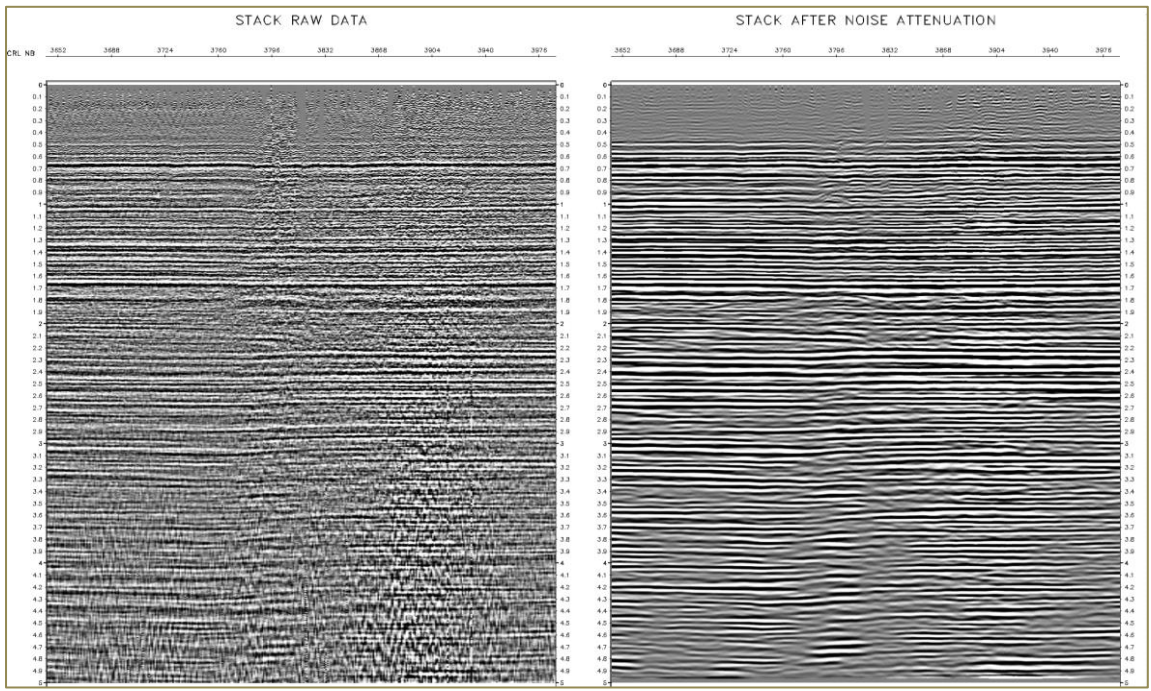

**Figure** 17 Brute stack before noise attenuation (b) Stack after noise attenuation

**Table 5** Summary of a noise attenuation processing sequence to current 2-D seismic data

| Sequential Processing Flow | Parameters Applied |
|---|---|
| Reformatting (**SEGIN**) | Anti-aliasing High Cut filer 100Hz/96dB/Oct |
| Geometry Up-date (Onset App.) (**ETQXY**) | Input of XY Coordinates into the seismic traces |
| Spherical divergence correction (**SDICO**) | Using P. Newman's formulae. |
| Random Noise attenuation (Despike) (**FDNAT**) (Shot domain) | High-amplitude noise (frequency-dependent and time-variant amplitude [threshold]). |
| Ground-Roll (GR) **AGORA** | group velocity 1000m/s |
| Radial trace Mixing (**RADMX**)(offset domains) | Assuming traces have NMO applied, stacking traces locally (offset domain) around each trace produces a noise suppressed version of the trace. |
| Structural Consistent Filtering (**SCIFL**) (offset domain) Non-stationary filter | Structurally consistent (SC) filter. A spatial window of 7 traces, Filter length (500 ms). Frequency range - above 12 Hz). |
| Dip-Dependent Median/trimmed mean filtering (**DDMED**) | NC (3), LWIN (300 ms), TI (0 ms), MAXDIP &MINDIP(6&-6 ms/trace), ND(5), SIGMA (0) |
| Time Variant Filter (**TVF**) | time (in ms)Filter Band (in Hz) 0-1100/ 14, 18-40, 50 Hz- 1400-5000 /12, 16-35, 45Hz |

5. **Conclusions**

The quality of seismic products and solutions being delivered today is, without doubt, a huge improvement compared to a few years ago. The rate of development, however, is very fast and it remains problematic for seismic data processing. 2-D seismic data acquired in the south of Iraq are presented here for noise attenuation .The processing was carried out in Iraq OEC's processing center. Testing was made at every step of noise attenuation; hence, the data is convoluted by a considerable amount of linear and impulse noise. The processing sequence was carried out with a maximum of true useful signals, and a wide range of modules was used for noise attenuation. We have tested the effectiveness of currently available modules and filters in our processing system for noise attenuation to improving 2-D seismic image quality. In this case of 2-D seismic data, the data set is processed with the processing sequence, including Frequency-Dependent Noise Attenuation (FDNAT) for random noise, the Adaptive groundroll Attenuation (AGORA) filters, Radial trace Mixing (RADMX), Non-stationary Structural Consistent Filtering (SCIFL) filter, Dip-Dependent Median/trimmed mean filtering (DDMED) and Time Variant Filter (TVF). FDNAT is determines the amplitude value in specific time window and after that compute the suitable threshold corresponding to the properties of the time windows and signal frequencies for detecting high energy amplitudes So, this filtering type offering fitting and steady threshold determination. And the results were effective in attenuation random noise in pre-stack data. Moreover, avoided signal data losses and increasing SNR. The adaptive (AGORA) filter is a data-driven approach we applied to resolving problem of coherent surface wave noise which offers advantages over conventional methods (modeling of surface and guided waves), so the data-driven estimate changes groundroll characteristics from the data, then the Adaptive modeling and estimate groundroll velocity, frequency, and aliasing. This filter is improving image quality without loss of data (reflection events). Also, the result was encouraging even if the seismic data set does not agree with the noise model and later it was very effective in increasing the SNR as we show in spectral analysis. RADMX suppressed coherent noise in the radial trace domain which takes advantage of the fact that linear noise separation from reflection events can be accomplished in R-T domain by aligning the transform coordinate trajectories with the coherent noise wavefronts in the *t-x* domain. This module is more valuable than some of the more traditional noise attenuation module. Because efficiently suppresses linear noise almost parallel to the long offset limbs of reflections. some very dispersive noises, such as a bit of the groundroll are very difficult to handle in the *f-k* domain but yield further easily in the R-T domain. The radial mixing process seems to be an influential way for attenuating steeply dipping noise which otherwise leaks into stack image. SCFIL module performs non-stationary filtering which attenuates random noise in common offset domain which is an effective solution to execute 2-D dip filtering, steered by spatially, local varying dip fields. This module by running dip-consistent manner Smoothing Structural in homogeneous areas while preserving trends, edges, and details, so, it give a best result in structural conformity. The results of (RADMX and SCFILL) are efficient and can reveal features and geological structures that were masked by noise present in the current data. DDMED applied a median or trimmed mean (non-linear) filter along the local dip of the data, thereby suppressing random noise and increasing trace-to-trace coherency and increasing the signal/noise ratio. This module is a robust, effective and relatively easy operation, and generally better at removing local high amplitude noise events than other techniques such as *f-k* and *f-x* filtering. So, it's realized that any type of coherent noise along a specific dip (e.g. ground roll) will be enhanced in the same manner. The last filter is TVF which it is often necessary final processing stage apply TVF to stacked data uncover the clearest possible image of stratigraphic boundaries. Over all, the final results shows a comprehensive noise attenuation processing sequence for current 2-D seismic data It is shown that an improved resolution and signal-to-noise ratio. Because encourage improvements in the final image quality in seismic section, these filtering procedure may support the interpreter in particular stratigraphic and structural analysis and identification of possible traps.

**Acknowledgments**

We would like to thank Oil Exploration Company (OEC) for their cooperation and allowing us to use CGG GEOVATION system and also, for permission to show the anonymous 2-D lines stacks and gathers that used to illustrate noise attenuation in this article. Special gratitude to Mr. Aws Riyadh and Mr. Hussain Ridha for their support.